\documentclass[11pt]{article}
\usepackage{amssymb,amsmath,amsfonts,palatino,amsthm}
\usepackage{graphicx}
\usepackage{enumerate}
\usepackage{amssymb}
\usepackage{url}                
\usepackage[footnotesize,ruled]{caption}

\usepackage{epstopdf}
\usepackage[parsep]{collref}

\newcommand{\beq}{\begin{equation}}
\newcommand{\eeq}{\end{equation}}
\newcommand{\bea}{\begin{eqnarray}}
\newcommand{\eea}{\end{eqnarray}}

\newcommand{\Mbh}{M_{\rm BH}}

\newcommand{\Mf}{M_d}

\newcommand{\MP}{M_{\rm Pl.}}

\newcommand{\LP}{\ell_{\rm Pl.}}
\newcommand{\lc}{\lambda_\mathrm{C}}
\newcommand{\Mm}{M_{\rm BH, min}}
\newcommand{\rgm}{r_{\rm g, min}}

\author{Jonas Mureika$^{1}$\thanks{\texttt{jmureika@lmu.edu
}}\ \ \& Piero Nicolini$^{2,3}$\thanks{\texttt{nicolini@fias.uni-frankfurt.de}}\\
$^{1}$Department of Physics, Loyola Marymount University, Los Angeles, CA, USA\\
$^{2}$Frankfurt Institute for Advances Studies, D-60438 Frankfurt, Germany\\
$^{3}$Institut f\"{u}r Theoretische Physik, 
J. W. Goethe-Universit\"{a}t, D-60438 Frankfurt, Germany\\
 }
\title{Self-completeness and spontaneous dimensional reduction}

\begin{document}
\date{\today}
\maketitle

\begin{abstract}
A viable quantum theory of gravity is one of the biggest challenges facing physicists.  We discuss the confluence of two highly expected features which might be instrumental in the quest of a finite and renormalizable quantum gravity -- spontaneous dimensional reduction and self-completeness.  The former suggests the spacetime background at the Planck scale may be effectively two-dimensional, while the latter implies a condition of maximal compression of matter by the formation of an event horizon for Planckian scattering. We generalize such a result to an arbitrary number of dimensions, and show that gravity in higher than four dimensions remains self-complete, but in lower dimensions it is not.  In such a way we established an ``exclusive disjunction'' or``exclusive or'' (XOR) between the
occurrence of self-completeness and dimensional reduction, with the goal of actually reducing the unknowns for the scenario of the physics at the Planck scale. Potential phenomenological implications of this result are considered by studying the case of a two-dimensional dilaton gravity model resulting from dimensional reduction of Einstein gravity.
\end{abstract}

\section{Introduction}
Unlike quantum descriptions of the other fundamental forces, gravitation is ill-behaved.  Most notably it is non-renormalizable, and as such no similar quantum description of gravity in four dimensions has been found.  String theory -- which is ultimately claimed to be the unifying theory of all fundamental interactions at the quantum level -- requires additional spatial dimensions, that, to date,
have not been verified experimentally. 
Alternatively the short distance behavior of the gravitational field might be improved by a mechanism called ``spontaneous dimensional reduction'', which opens the route towards an effective two-dimensional renormalizable formulation of quantum gravity \cite{AJL05,LaR05,Ben09,Mod09,NiS11,LaC11,Cal12a,Cal12b,MoN10a,Car09,Car11,CaG11,Mur12}.  A complementary character of gravity at the Planck scale is ``self-completeness'', namely the emergence of a natural cut-off that masks spacetime pathologies in the ultraviolet regime. While each of them alone might prove to be instrumental in the formulation of a consistent quantum theory of gravity, for the first time we explore the scenario emerging from the simultaneous presence of both effects by critically reviewing currents beliefs on the subject.

\section{Self-completeness}
In the standard picture of quantum field theory, shorter length scales of a physical system become visible as one increases the energy of the probe. The Compton wavelength -- representing the best possible resolution of the position of a particle -- is governed by the well-known relation 
\begin{equation}
\lambda_{\rm C}=\frac{2\pi\hbar}{Mc}~~,
\end{equation}
where $M$ is the mass associated to the particle under consideration.   The current LHC working energy $\sim 8 \mathrm{TeV}$ thus corresponds to matter compressed  within the exceedingly minuscule distance of $\sim 10^{-19}\ \mathrm{m}$ \cite{LHC12a}. In principle one may be tempted to conclude that $\lambda_{\rm C}$ can be arbitrarily small, provided enough energy can be supplied to a particle.   By probing shorter and shorter distances, however, one enters a regime in which the background spacetime manifold becomes significantly disturbed by the energy involved in the process.  Such a disturbance prevents the localization to better accuracies than a fundamental (minimal) length scale $\ell$.  

In other words, gravity prevents the compression of matter beyond certain distances due to the formation of a black hole.  When
\begin{equation}
\lambda_\mathrm{C}\sim r_{\rm g}~~~,
\end{equation}
the particle collapses to form a black hole and cannot be made smaller. Here $r_{\rm g}=2GM/c^2$ is the gravitational radius associated to the particle mass $M$.   Unlike in the case of the Compton wavelength, a further increase of the energy of the system will result in a bigger gravitational radius.   As a result the above condition sets the minimum attainable size when both gravity and quantum mechanics are concerned.  Not surprisingly, one finds
\begin{eqnarray}
\ell\equiv r_{\rm g, min}=2\sqrt{\pi}\LP
\end{eqnarray}
and correspondingly
\begin{eqnarray}
M_{\rm BH, min}=\sqrt{\pi}\MP 
\end{eqnarray}
where $\LP\simeq 1.6\times 10^{-35}\ \mathrm{m}$ and $\MP\simeq 1.22 \times 10^{16}\ \mathrm{TeV}/c^2$ are the Planck length and the Planck mass, respectively (see left plot in Fig \ref{fig1}).

The above relations show that gravity is {\it self-complete}, namely it protects the ultraviolet regime by setting quantum mechanical limits to length and energy.  This result is often employed as an elegant argument to downplay the problem of spacetime curvature singularities, which would always be inaccessible to external probes due to the presence of an event horizon \cite{AuS04,Adl10,DvG10,DFG11}. This viewpoint has been reinforced by a general result obtained in an independent way by several quantum gravity modifications of black hole metrics \cite{BoR00,Nic05,NSS06b,Mod06,Mod08,SSN09,Nic09,NiS10,MMN11,CMP11,Mod11,Mod12,Nic12}: even at the terminal stage of the evaporation the probe of the shortest scales in the vicinity of the curvature singularities would not be possible for the formation of zero temperature black hole remnants, \textit{i.e.}, extremal configurations occurring also in the case of non-rotating, neutral black holes \cite{SpA11,SpS12}. 

We now demonstrate the robustness of this line of reasoning by its extension to the case of additional spatial dimensions, a usual requirement when considering  possible quantum gravity phenomenology at the TeV energy scale. In a $(d+1)$-dimensional spacetime one finds
\begin{equation}
\ell\equiv\rgm=\left(\frac{4\pi\hbar G_d}{c^3}\right)^{\frac{1}{d-1}}
\label{eq:ell_d}
\end{equation}
\begin{equation}
\Mm = 2^{\frac{d-3}{d-1}}\left(\pi\hbar\right)^{\frac{d-2}{d-1}} c^{-\frac{d-4}{d-1}} G_d^{-\frac{1}{d-1}}
\label{eq:mass_d}
\end{equation}
Here $G_d$, the $d$-dimensional gravitational constant, reads
\begin{equation}
G_d=2K_d  \ \pi^{1-\frac{d}{2}}\Gamma\left(\frac{d}{2}\right)\frac{c^3}{\hbar}\left(\frac{\hbar}{\Mf c}\right)^{d-1}
\label{eq:G_d}
\end{equation}  
where $\Mf\sim 1 \mathrm{TeV}/c^2$ is the $d$-dimensional Planck mass and $K_d$ is a constant that varies according to the definitions of $\Mf$. The only requirement for $K_d$ is a matching between $G_3$ and Newton's constant $G$, i.e., $K_3=1$ for $M_3=\MP$. The other factors in (\ref{eq:G_d}) come for the Gauss law in $d$ dimensions. For the present discussion and without loss of generality we can simply set $K_d=1$ for all $d$ as in \cite{MaM11}. As a result the limits for energy and length turn out to be
\begin{eqnarray}
\ell\equiv r_{\rm g, min}=8^{\frac{1}{d-1}}\pi^{-\frac{1}{2}\left(\frac{d-4}{d-1}\right)} \left[\Gamma\left(\frac{d}{2}\right)\right]^{\frac{1}{d-1}}\left(\frac{\hbar}{\Mf c}\right)
\end{eqnarray}
and 
\begin{eqnarray}
M_{\rm BH, min}=2^{\frac{d-4}{d-1}}\pi^{\frac{3}{2}\left(\frac{d-2}{d-1}\right)} \left[\Gamma\left(\frac{d}{2}\right)\right]^{-\frac{1}{d-1}}\Mf .
\end{eqnarray}
By varying $d=4-10$ we find $\ell=(1.23-2.00)\hbar/M_d c$ and $\Mm=(\pi-5.13)M_d$, indicating that microscopic black holes are at the reach of the typical LHC energies (see \cite{MNS12,CMO12} for the latest production rate estimates).

\section{Spontaneuous dimensional reduction}
In the above derivation we did not invoke any specific quantum gravity character other than the fact that the energy involved are of the order of the Planck scale. The issue arises since the spacetime at  this length scale might be radically different than its conventional picture.  Due to its intrinsic graininess, the quantum spacetime has often been described in terms of a fractal surface which smoothly approaches the structure of differential manifold only in the infrared limit. 
As a result, the very concept of spacetime dimension becomes ill defined or at the very least needs revising.   A reliable indicator of the dimension of a fractal is given by the spectral dimension, \textit{i.e.}, the actual dimension perceived by a random walker. In several approaches to quantum gravity the spectral dimension has a general form like  
\begin{equation}
\mathbb{D}=D_{\rm IR}-\frac{a}{b+(\sigma/\ell^2)}
\end{equation}
where the infrared dimension is $D_{\rm IR}=d_{\rm IR}+1$, $\sigma$ is the diffusion time with the dimensions of a length squared, $a$ and $b$ are dimensionless constant depending on the specific model of quantum manifold under consideration \cite{AJL05,LaR05,Ben09,Mod09,NiS11,LaC11,Cal12a,Cal12b,MoN10a,Mod11,Mod12}. While at large diffusion times, \textit{i.e.}, $\sigma/\ell\gg a, b$, the spectral dimension $\mathbb{D}=D_{\rm IR}$ as expected, in the opposite regime, \textit{i.e.}, $\sigma/\ell\ll a, b$, the spectral dimension decreases to the ultraviolet value $D_{\rm UV}=D_{\rm IR}-a/b$. Such a behavior is connected to the potential power counting renormalizable character of gravity which in two dimensions exhibits a dimensionless coupling constant, as is evident from (\ref{eq:G_d}) for $d=1$. 

\begin{figure}[!ht]

  \centering
 \includegraphics[height=7cm]{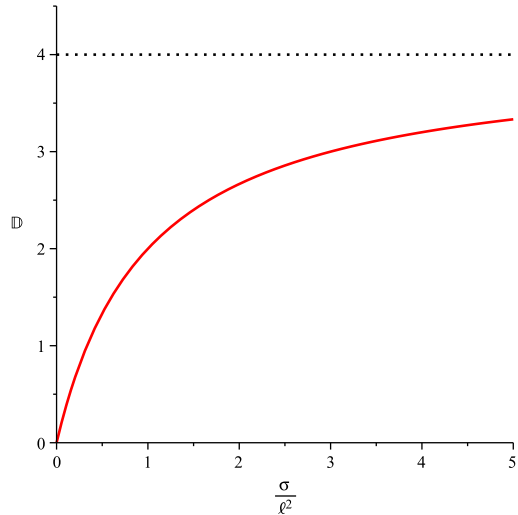}

 \caption{\label{fig3}Example of dimensional reduction \cite{MoN10a}: the spectral dimension $\mathbb{D}$ as a function of the diffusion time $\sigma$ for $D_\mathrm{IR}=4$. A decrease of spacetime dimension already occurs in the sub-Planckian regime, \textit{i.e.}, $\sigma>\ell^2$,  in order to match the value $\mathbb{D}=2$ for $\sigma=\ell^2$. As a consequence semiclassical arguments are still valid for $2<\mathbb{D}<D_\mathrm{IR}$. }
   
  \end{figure}

The desired reduction to a two-dimensional spacetime at the Plank scale fixes the range of the above parameters, \textit{i.e.}, $a=(D_{\rm IR}-2)(b+1)$. Another condition can be derived from the requirement $\mathbb{D}\geq 0$: even if negative spectral dimensions are admissible for fractals, for most models of quantum gravity in the literature \cite{AJL05,LaR05,Ben09,Mod09,NiS11,LaC11,Cal12a,Cal12b} the spectral dimension turns out to be strictly positive.   As a result,  one obtains that $b\geq \frac{1}{2}D_{\rm IR}-1$.
Figure~\ref{fig3} demonstrates an example of dimensional reduction compatible with the above constraints (for details about its derivation in the framework of noncommutative geometry see \cite{MoN10a}).
In summary, if the quantum spacetime underwent a spontaneous dimensional reduction to $D_{\rm UV}=2$ ($d_{\rm UV}=1$), the non-renormalizability of gravity would just be an ``apparent'' low energy, classical effect \cite{tHo93}. 

Given these ingredients, the usual argument for gravity self-completeness has to be reviewed. Eq. (\ref{eq:ell_d}) and (\ref{eq:mass_d}) are singular for $d=1$. We therefore have to reconsider the behavior of $r_{\rm g}$ in lower dimensions. Before doing this we see that the gravitational potential due to the integration of the Gauss law in $(1+1)$-dimensions reads
\begin{equation}
\phi_1(x)=-2\pi \frac{c^3}{\hbar} M x + \phi_0~~,
\label{potential1d}
\end{equation}
where $x$ is the spatial coordinate and $\phi_0$ is an integration constant.  The above relation shows that the quantity $G_1M/c^2$ no longer has the dimension of a length, but rather inverse length. This implies that the gravitational radius  $r_{\rm g, 1}$ in $(1+1)$-dimensions will have a intriguing new behavior, {\it i.e.}
\begin{equation}
r_{\rm g, 1}\propto \lc .
\end{equation}
The above result implies that in case of spontaneous dimensional reduction at the Planck scale the conventional arguments in support of gravity self-completeness is no longer valid. 
Such a conclusion holds for any the profile of the metric coefficients in $(1+1)$-dimensions, because the gravitational coupling $G_1$ becomes dimensionless and actually there is no Planck scale to discriminate between classical and quantum black holes.  In addition, this fact allows one to safely employ a semiclassical model of dimensionally reduced spacetime to illustrate the associated phenomenology.

\section{Dilaton gravity}
A more detailed argument for this case may be made by considering a specific (relativistic) model of $(1+1)$-dimensional gravity able to circumvent the triviality of Einstein equations. This is usually achieved by the so called dilaton gravity (DG) models in which the dilaton, an extra field dated back to Kaluza-Klein theories and reappeared in the context of string theory, accounts for some key features of the higher dimensional theory (for a review see \cite{GKV02}).   A generic action for $(1+1)$-dimensional dilaton gravity is 
\beq
S_{\rm DG} = \frac{c^4}{8\pi G_1} \int d^2x \sqrt{-g} \left[ \psi R + U(\psi) (\nabla \psi)^2 - 2 V(\psi) \right]~~
\label{generic2d}
\eeq
where $\psi$ is the dilaton field and the functions $U(\psi),V(\psi)$ are model-dependent potentials
(see \cite{GrM06} for a comprehensive tabular summary). 
For the purposes of the present discussion, however, we invoke
the mechanism of spontaneous dimensional reduction, $d_{\rm IR}\to d_{\rm UV}= 1$, to determine the profile of the above potentials even if, as already stated, the choice will not restrict the broadness of our conclusions. Starting from the $(d_{\rm IR}+1)$-dimensional action for general relativity
\begin{equation}
S_{(d_{\rm IR}+1)}=\int d^{(d_{\rm IR}+1)}x \sqrt{-g}\left(\frac{1}{\kappa_d}R+{\cal L}_{\rm m}^{(d_{\rm IR}+1)}\right)
\label{eq:d_action}
\end{equation}
where ${\cal L}_{\rm m}^{(d_{\rm IR}+1)}$ is the matter Lagrangian, one can show that, by expanding (\ref{eq:d_action}) in powers of $(d_{\rm IR}-1)$, the theory reduces to 
\beq
S_{(1+1)} = \int d^2x~\sqrt{-g} \left[\left(\frac{c^4}{8\pi G_1} \psi R-\frac{1}{2}(\nabla \psi)^2\right) +{\cal L}_{\rm m}^{(1+1)}\right],
\label{eq:robb_action}
\eeq
in the limit $d_{\rm IR}\to 1$ provided that $\kappa_d=4\pi(1-d_{\rm IR})G_{\rm IR}/c^4$ \cite{MaR93}.  A more general mechanism of dimensional reduction from Einstein gravity to an effective $(1+1)$-Liouville gravity have been studied, with the addition of extra terms in the action \cite{Jac06,GrJ10}
\bea
&&\hspace{-0.4cm} S =  \int\; d^2x \sqrt{-g} \left[ \frac{c^4}{8\pi G_1}\psi R - \frac{1}{2}(\nabla \psi)^2 - C e^{-2\psi} +{\cal L}_{\rm m}^{(1+1)}\right]\nonumber\\ &&+ \left({\rm boundary~terms}\right)
\label{eg:grumiller_action}
\eea
where $C$ is a constant.  This enhanced model contains (\ref{eq:robb_action}) as a subset and provides a robust match for dimensionally-reduced spherical gravity.

\begin{figure}[!ht]
  \centering
 \includegraphics[height=7cm]{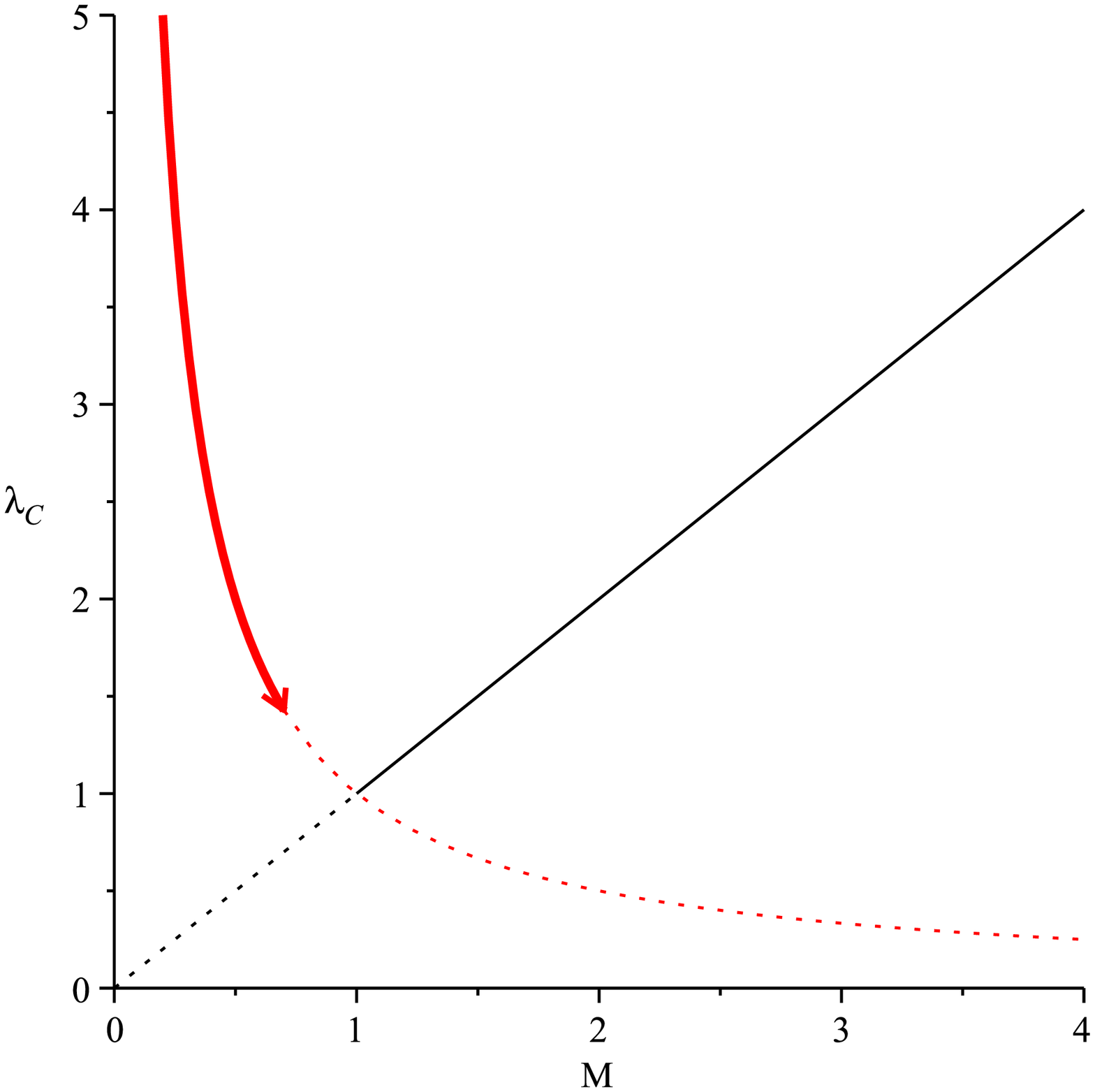}
 \includegraphics[height=7cm]{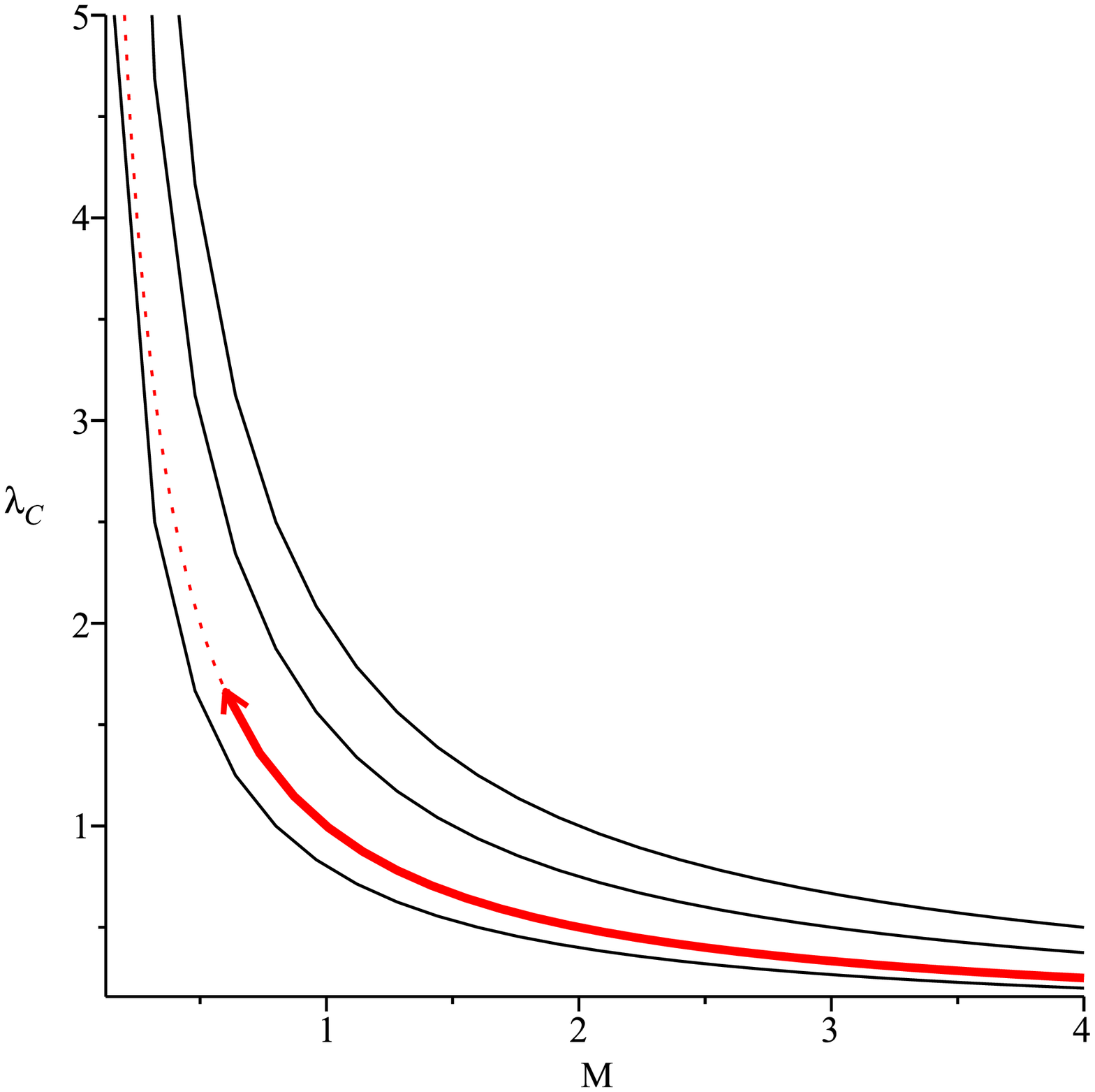}
 \caption{\label{fig1}The scale-size in the infrared regime (left) as a function of the particle mass $M$, including the Compton wavelength $\lambda_{\rm C}$ (solid thick hyperbola) and the gravitational radius $r_{\rm g}$ (solid linear) for classical, trans-Planckian black holes. The intersection of the above curves gives the minimum mass for forming a black hole as well as the minimum detectable length scale as it appears in the infrared limit. The direction of the arrow shows the increase of mass from sub-Planckian to Planckian scales. (Right) Scale-size in the ultraviolet regime as a function of $M$, where the effective spacetime dimension is 2 and gravitational radii are hyperbolae (thin solid curves).  The thick solid curve represents a quantum black hole, \textit{i.e.}, a black hole which is at the same time an elementary particle.  Classical as well quantum black holes can occur for any mass.  By Hawking emission ``trans-Planckian'' black holes decay into ``sub-Planckian'' black holes according to the direction of the arrow. 
 In both plots we adopted the units $\ell=1$, $M_{\rm BH, min}\sim M_{d_{\rm IR}}=1$ }
  \end{figure}

The above theories provide a faithful description of gravity in $(1+1)$-dimensions: they 
preserve classical and semiclassical properties of higher dimensional gravity \cite{MST90,SiM91,MMS91,MoM91,MaS92} and can be simply connected to other dimensional reduced gravity proposals \cite{Jac84,CaJ92,Man94}. 
%
By solving the corresponding field equations one finds the metric
\beq
ds_1^2 = -\left(\ \frac{2 G_1 M}{c^2} |x| -C\ \right) dt^2 + \frac{dx^2}{\left( \ \frac{2 G_1 M}{c^2}|x| -C\ \right)} \label{g11}
\eeq
where $x$ is the spatial coordinate and 
the constant $C$,  depending on its sign,  correlates to specific physical objects in our results. 
The horizon is thus
\beq
r_{\rm g, 1}\equiv |x|_{\rm H}=\frac{c^2}{2G_1 \Mbh} C\label{r11}
\eeq
which occurs only for $C>0$. 
In this case, the usual self-completeness ``condition'' results in the mass-independent expression fixing the constant $C$,
\beq
\frac{c^2}{2G_1\Mbh}C = \frac{2\pi\hbar}{\Mbh c}~~~\Longrightarrow~~~ C = \frac{4\pi\hbar}{c^3}G_1 ~~~\Longrightarrow ~~~ C = 8\pi^2 .
\eeq
This is a striking result, in that it demonstrates the self-{\it incompleteness} of gravitation in a two-dimensional spacetime (see right plot in Fig. \ref{fig1}).  In the $(1+1)$-dimensional regime, the algebraic form of the defining classical quantity (the horizon radius) is equivalent to the defining quantum characteristic (the Compton wavelength).  This suggests that there is no classical-quantum boundary in two-dimensional spacetime, and hence gravitation is naturally quantum mechanical at this scale.

The value and sign of $C$ have the following repercussions. First, for $C\neq 8\pi^2$ one has ``classical black holes'', \textit{i.e.}, black holes that do not meet the condition for being elementary particles. Second, in a hypothetical ``trans-Planckian'' collision  (see note 
\footnote{In this case, the terms ``sub- and trans-Planckian'' follow a colloquial definition as  measured by an external observer that supplies the energy for the collision. Since $G_1$ is a dimensionless quantity, we should not be allowed to define a Planck mass or equivalently a Planck length.  
We have retained the conventional definition by using quotation marks, however, since we have distinguished between the spectral dimension -- the actual dimension ``perceived'' by probes ({\it e.g.} particles) -- and the conventional topological dimension, {\it i.e.} $d_{\rm IR}$. \label{note}}   
for the use of quotation marks of ``(sub/trans)-Planckian'' in such a context), the particle Compton wave length can be made smaller than $\hbar/M_{d_{\rm IR}} c$, the $d_{\rm IR}$-dimensional Planck length, resulting in the formation of tiny, non-classical, ``trans-Planckian'' black holes (i.e. $\Mbh>M_{d_{\rm IR}}$) for $C=8\pi^2$.   Third, when $C < 0$, the horizon is undefined and the object is an elementary particle.

Lastly, we emphasize there is no minimum energy scale that defines a black hole. This is reflected by the fact that for a given mass $\Mbh$, it is always possible to have black hole smaller (for $0<C<8\pi^2$) or larger (for $C>8\pi^2$) of the corresponding Compton wavelength of a particle of the same mass.  Note this is also true in the case of the extension of the present theory to the case of a $(1+1)$-dimensional, singularity free, non-commutative geometry: no extremal black hole configurations occur \cite{MuN11}. Unlike in the four-dimensional (or higher) case, ``sub-Planckian'' black holes ({\it i.e.} $\Mbh<M_{d_{\rm IR}}$)  are possible in $(1+1)$-dimensional spacetime.  They cannot, however, be the direct product of a dimensional reduction mechanism (which requires Planckian energies), but rather a transient state between the formation of ``Planckian'' (or ``trans-Planckian'') $(1+1)$-black holes and their complete evaporation or some other stable configuration.


We can see this by studying the thermal properties of these dimensionally reduced black holes. From the periodicity of the Euclidean time of (\ref{g11}) one finds the temperature
 \begin{equation}
T=\frac{\hbar G_1}{2\pi k_{\rm B}c}M_{\rm BH}
\end{equation}
which becomes $T=M_{\rm BH}c^2/k_{\rm B}$ by using the definition of $G_1$. We see that heavier (``trans-Planckian'') holes are smaller and hotter. However the positive sign of the heat capacity ${\cal C}\equiv \partial (Mc^2)/\partial T=k_{\rm B}$ indicates that the evaporation slows down as the black hole loses energy, the contrary of what happens in the four (and higher) dimensional case. This fact results in a peculiar form of the Stefan-Boltzmann law, which in $(1+1)$-dimension reads \cite{Mur12}
\begin{equation}
-\frac{dM_{\rm BH}}{dt}\propto T^2\propto M_{\rm BH}^2.
\label{eq:sb}
\end{equation}
From the above relation we see that the relative mass loss $\frac{1}{M_{\rm BH}}|\frac{\partial M_{\rm BH}}{\partial t}|\sim M_{\rm BH}$ increases with the mass. Lighter black holes are bigger, colder and tend to evaporate at a slower rate. We have to restrict our analysis to mass parameters compatible with the case under 
consideration, however, \textit{i.e.}, initial ``Planckian'' (or ``trans-Planckian'') black holes with  mass $M_{\rm BH, TP}\gtrsim M_{d_{\rm IR}}$ decaying into smaller ``sub-Planckian'' holes with mass $M_{\rm BH, SP}< M_{d_{\rm IR}}$.  Consequently, we can estimate evaporation times $t_{\rm ev}$ for such a decay by integrating (\ref{eq:sb}) within the corresponding mass interval. As a result we have that
\begin{equation}
t_{\rm ev}\sim\left(\frac{M_{d_{\rm IR}}}{M_{\rm BH, SP}}-\frac{M_{d_{\rm IR}}}{M_{\rm BH, TP}}\right)\left(\frac{\hbar}{M_{d_{\rm IR}} c^2}\right), 
\end{equation}
which, for $M_{\rm BH, SP}\lesssim M_{d_{\rm IR}}$, sets the typical time scale associated with $M_{d_{\rm IR}}$, {\it i.e.} $t_{\rm IR}\equiv\left(\frac{\hbar}{M_{d_{\rm IR}} c^2}\right)$. Smaller holes would have $t_{\rm ev}>t_{\rm IR}$, resulting into quasi-stable objects.

To better understand the fate of such ``sub-Planckian'' black holes with mass $M_{\rm BH, SP}\lesssim M_{d_{\rm IR}}$  we should go back to the mechanisms of dimensional reduction, {\it i.e.} the phase in which (trans-) ``Planckian'' black holes forms. The process is again regulated by the time scale $t_{\rm IR}$. The effective spacetime dimension smoothly decreases from $D_{\rm IR}$ to 2  in a time lapse of the order $\sim t_{\rm IR}$  during which the system undergoes a temporary $(2+1)$-dimensional regime. This corresponds to having Newton's potential of the form
\begin{equation}
\phi_2=\phi_0-G_2 M\ln\left(r/\LP\right)=\phi_0-2c^2\left(\frac{M}{\MP}\right)\ln\left(r/\LP\right)
\label{eq:2+1_Newton}
\end{equation}
where we have used the relation $G_2=2(c^3/\hbar)(\hbar/\MP c)$ and $\phi_0$ is another integration constant (distinct from that in (\ref{potential1d})). From (\ref{eq:2+1_Newton}) we see that the $(2+1)$-dimensional spacetime is a special case in which $G_2M/c^2$ is a dimensionless quantity, signaling the absence of event horizons. This fact is in agreement with black holes derived from the BTZ action, that can only exist in anti-deSitter universes \cite{BTZ92}.   In addition Newton's potential is divergent both in the UV and IR regimes, a borderline situation between IR-dimensional and UV-dimensional cases. As a consequence we interpret the $(2+1)$-dimensional case as the regime of a phase transition between the higher-dimensional and the two-dimensional black holes. 

In light of this reasoning we can argue that ``sub-Planckian'' black holes would not completely evaporate,  but they would rather undergo to a ``dimensional oxidation'', namely the opposite process to the dimensional reduction. In other words the initial (trans-) ``Planckian'' black holes would decay into transient ``sub-Planckian'' black holes in a time $t_{\rm ev}\sim t_{\rm IR}$ to oxidate into ordinary sub-Planckian particles in the $(d_{\rm IR}+1)$-dimensional spacetime. This might change the conventional signatures for black hole detection in particle accelerators \cite{CaL10} and explain the reason why latest data tend to exclude their observation \cite{CMS11b,CMS12}.

\section{Conclusions}
In summary, we have presented in this paper new insights on the nature of lower-dimensional gravitation, specifically that the well-known self-completeness condition does not hold in spacetimes below $(3+1)$-dimensions.  The immediate consequence of this result is that the mass of a quantum black holes formed in a $(1+1)$-dimensional Universe is unbounded from below.   Consequently, this would suggest that evaporating black holes will eventually reach a new non-thermal phase in their evolution if Planckian dimensional reduction theories are correct.  For primordial black holes of the appropriate masses, this has already occurred.  Such objects can thus be considered a new catalyst in early Universe physics such as formation mechanism, a possible new candidate for dark matter, or even a new bi-product of high-energy collisions at the specified energy.   Depending on the scale of such transitions, physical evidence could be just within reach of present or future collider experiments, ultra-high energy cosmic ray detectors, of other cosmological probes.  If detected, signals of dimensional reduction would present a profound confirmation of these innovative formulations of spacetime structure, and would revolutionize our understanding of the Universe.

Although the above scenario is intriguing, we do not wish to further speculate without a corroboration from a more detailed analysis.  In this paper, we safely prefer not to assume an \textit{a priori} position but rather we simply conclude that self-completeness and dimensional reduction are conflicting occurrences, connected by an ``exclusive disjunction'' or ``exclusive or'' (XOR). This logic is instrumental for reducing the variety of possible approaches in the description of physics at the Planck scale. 

\subsection*{Acknowledgments}
JM would like to thank the generous hospitality of the Frankfurt Institute for Advanced Studies, at which this work was done. This work has been supported by the project ``Evaporation of microscopic black holes'' (PN) of the German Research Foundation (DFG), by the Helmholtz International Center for FAIR within the framework of the LOEWE program (Landesoffensive zur Entwicklung Wissenschaftlich-\"{O}konomischer Exzellenz) launched by the State of Hesse (PN), partially by the European Cooperation in Science and Technology (COST) action MP0905 ``Black Holes in a Violent Universe'' (PN) and also by a Continuing Faculty Grant from Loyola Marymount University (JM). The authors thank Gianluca Calcagni, Daniel Grumiller, Robert Mann, Euro Spallucci for valuable comments on an earlier version of the manuscript.



\providecommand{\href}[2]{#2}\begingroup\raggedright\endgroup

\end{document}